\documentstyle[12pt,epsf]{article}

\begin{document}

\title{\vspace{-1 in}Kinetics of Catalysis with Surface Disorder}

\author{D.A.Head and G.J.Rodgers}

\date{Department of Physics, Brunel University, Uxbridge,\\
Middlesex. UB8 3PH, UK}

\maketitle

\begin{abstract}
We study the effects of generalised surface disorder on the
mon\-omer-monomer model of heterogeneous catalysis, where disorder
is implemented by allowing different adsorption rates for each lattice site.
By mapping the system in the reaction-controlled limit onto a kinetic
Ising model, we derive the rate equations for the one and two-spin
correlation functions. There is good agreement between these equations and
numerical simulations.
We then study the inclusion of desorption of monomers from the substrate,
first by both species
and then by just one, and find exact time-dependent solutions
for the one-spin correlation functions.  
\end{abstract}


\vspace{0.6 in}
e-mail: David.Head@brunel.ac.uk, G.J.Rodgers@brunel.ac.uk
\vspace{0.15 in}

\noindent{\today}

\newpage

\noindent{\bf I. INTRODUCTION}
\bigskip

Diffusionless surface-reaction models
were first introduced by Ziff, Gulari and Barshad~\cite{ziff2},
who investigated a monomer-dimer reaction corresponding to the
chemical reaction $2CO+O_{2}\rightarrow 2CO_{2}$ on a catalytic surface.
A well-studied variant~\cite{ziff3,meakin,evans} employs
the simpler monomer-monomer reaction, described by

\begin{eqnarray}
A_{gas}+S & \stackrel{\rm k_{A}}{\longrightarrow} & A_{surface} \nonumber\\
B_{gas}+S & \stackrel{\rm k_{B}}{\longrightarrow} & B_{surface} \nonumber\\
A_{surface}+B_{surface} & \stackrel{\rm k_{R}}{\longrightarrow} & AB_{gas}+2S,
\end{eqnarray}

\noindent{where
S denotes an empty site.
This process exhibits a {\em kinetic phase}
when there are equal propensities of A and B species, in which the
long-time kinetics become dominated by domain coarsening.
Mean-field analysis~\cite{ben-A}, in which every site is taken to be
connected to every other site in a `complete graph', demonstrated that
finite lattices will always saturate - that is, the lattice will
either become full of  A's, or full of B's, and the process will stop.
Krapivsky~\cite{krapivsky} recently solved the model exactly in the
reaction-controlled limit $k_{R}\rightarrow\infty$ by mapping
the system onto the standard Ising model.
}

Many enhancements to these models have been studied with a view to
more closely modelling actual chemical processes, including
nearest neighbour excluded adsorption~\cite{zhuo} and
surface diffusion~\cite{evans,vlachos}. However, only
recently have the effects of surface disorder been touched
upon by Frachebourg\linebreak {\em et.al.}~\cite{redner}.
They chose to model a disordered surface by taking a lattice
of two different types of site,
one which favours adsorption by the A-species, and one which favours
adsorption by the B's. They showed numerically
that such disorder allows for a reactive equilibrium in two
dimensions.

In this paper, we extend the analytical method used in~\cite{krapivsky}
to a general
form of surface disorder, based on~\cite{redner} but allowing for a range
of different types of site in the lattice. Furthermore, we also investigate
separately the effects
of desorption in the system. All the results presented
are for the physically relevant case of two dimensions.

This paper is organised as follows. In section~II we define the model
and derive
the general rate equations for the n-spin correlation functions.
In section~III,
these equations are applied to a model similar to that in~\cite{redner} and
their solutions are
compared to numerical simulations. In sections~IV and V we include the effects
of desorption, first by both species and then by just one, and derive
exact solutions. The conclusions are summarised in section~VI. 

\bigskip
\bigskip
\noindent{\bf II. RATE EQUATIONS}
\bigskip

We consider the surface reaction $A+B\rightarrow2S$ on a periodic $L\times L$
square lattice, ignoring the effects of diffusion and desorption.
For simplicity, we take the
reaction-controlled limit, where the adsorption of A and B
species is taken to be infinitely fast so that the substrate
is always full. The algorithm employed here is to select a nearest-neighbour
(NN) pair at random, check for an AB-reaction, and, if so, remove
the particles and immediately refill both sites. 

With the usual homogeneous model, the probability of filling a site with
and A or B is independant of the site chosen - in this model, however,
that probability is allowed to vary. Specifically, we introduce the
{\em site inhomogeneity} matrix $P_{ij}$, $0\leq P_{ij}\leq 1\; \forall i,j$,
such that the probability of filling the site~$(i,j)$ with
an A is given by $P_{ij}$ (or, equivalently,
a probability $1-P_{ij}$ of filling the site with a B).

Since in the reaction-controlled limit each site $(i,j)$
has only two possible states,
we can map this model onto an Ising model with mixed Glauber-Kawasaki
dynamics~\cite{krapivsky},
identifying A's with $S_{ij}=+1$ and B's with $S_{ij}= -1$.
The master equation for $P(S,t)$, the probability distribution for
the system to be in the state $S=\{S_{ij}\}$ at time t, is

\begin{eqnarray}
\frac{d}{dt}P(S,t)&=&\sum_{i,j}[\: U_{ij}(F_{ij}S)P(F_{ij}S,t)-U_{ij}(S)P(S,t)
\: ]\nonumber \\
&+&\sum_{i,j}[\: V_{ij}(F_{ij}F_{i+1j}S)P(F_{ij}F_{i+1j}S,t)-V_{ij}(S)P(S,t)
\: ]\nonumber \\
&+&\sum_{i,j}[\: W_{ij}(F_{ij}F_{ij+1}S)P(F_{ij}F_{ij+1}S,t)-W_{ij}(S)P(S,t)
\:].
\label{k:master}
\end{eqnarray}

The flip operator~$F_{ij}$ acts on the system state vector $S$
by flipping the sign of the $S_{ij}$ component, leaving the
remaining components unchanged.
$U_{ij}$ corresponds to Glauber spin-flip dynamics~\cite{glauber},
whereas $V_{ij}$ and $W_{ij}$
correspond to Kawasaki exchange dynamics. Equation~(\ref{k:master})
is identical
the the homogeneous case, except that now the full
expressions for $U_{ij}$,$V_{ij}$ and $W_{ij}$ are given by
 
\begin{eqnarray}
\label{uij}
4\tau_{1}U_{ij} & = &
(1-S_{ij}S_{i+1j})\{1-d_{ij}+S_{ij}(1-a^{+}_{ij})\} \nonumber\\
&&+\; (1-S_{ij}S_{i-1j})\{1-d_{i-1j}+S_{ij}(1-a^{+}_{i-1j})\}
\nonumber\\
&&+\;(1-S_{ij}S_{ij+1})\{1-e_{ij}+S_{ij}(1-b^{+}_{ij})\}\nonumber\\
&&+\;(1-S_{ij}S_{ij-1})\{1-e_{ij-1}+S_{ij}(1-b^{+}_{ij-1})\}, \\
\label{vij}
4\tau_{2}V_{ij}&=&(1-S_{ij}S_{i+1j})\left\{d_{ij}+a^{-}_{ij}
S_{ij}\right\}, \\
\label{wij}
4\tau_{2}W_{ij}&=&(1-S_{ij}S_{ij+1})\left\{
e_{ij}+b^{-}_{ij}S_{ij}\right\},
\end{eqnarray}

\begin{noindent}where the constant coefficients $a_{ij}^{\pm}$,$b_{ij}^{\pm}$,
$d_{ij}$ and $e_{ij}$ are related to the inhomogeneity matrix $P_{ij}$,
\end{noindent}

\begin{eqnarray}
a^{\pm}_{ij} & = & P_{i+1j}\pm P_{ij}, \nonumber \\
b^{\pm}_{ij} & = & P_{ij+1}\pm P_{ij}, \nonumber \\
d_{ij} & = & P_{ij}+P_{i+1j}-2P_{ij}P_{i+1j}, \nonumber \\
e_{ij} & = & P_{ij}+P_{ij+1}-2P_{ij}P_{ij+1}.
\end{eqnarray}

We proceed by deriving the rate equations for the one and two-spin
correlation functions, where the general n-spin function
is given by

\begin{equation}
\langle S_{i_{1}j_{1}}\ldots S_{i_{n}j_{n}}\rangle
=\sum_{S}S_{i_{1}j_{1}}\ldots S_{i_{n}j_{n}}P(S,t).
\label{spincor}
\end{equation}

Using this and~(\ref{k:master}), some lengthy but straightforward
calculations result in the following hierarchy of differential equations,
using the renormalised time scale $\tau$ defined by 
$\tau^{-1} = \tau_{1}^{-1}+\tau_{2}^{-1}$, and setting
$\tau_{1}=\tau_{2}$,

\begin{eqnarray}
\label{1pt}
4\tau \frac{d}{dt}\langle S_{ij}\rangle&=&\Delta_{ij}\langle S_{ij}\rangle
+(1-2P_{ij})\langle S_{ij}\{\Delta_{ij}S_{ij}\}\rangle, \\
4\tau \frac{d}{dt}\langle S_{ij}S_{kl}\rangle& = & (\Delta_{ij}+\Delta_{kl})
\langle S_{ij}S_{kl}\rangle \nonumber\\
&&+(1-2P_{ij})\langle S_{ij}S_{kl}\{\Delta_{ij}S_{ij}\}\rangle\nonumber\\
&&+(1-2P_{kl})\langle S_{ij}S_{kl}\{\Delta_{kl}S_{kl}\}\rangle .
\end{eqnarray}

\vspace{-0.10 in}\hfill\ldots for $|i-k|+|j-l|>1$
\bigskip

Here, $\Delta_{ij}\langle S_{ij}
\rangle =-4\langle S_{ij}\rangle + \langle S_{i+1j}\rangle +
\langle S_{i-1j}\rangle +\langle S_{ij+1}\rangle +\langle S_{ij-1}
\rangle $ is the discrete Laplacian.
For $|i-k|+|j-l|=1$, i.e. for nearest neighbour 2-point correlations, the
rate equation has a more complex form. For example,

\newpage

\begin{eqnarray}
4\tau \frac{d}{dt}\langle S_{ij}S_{i+1j}\rangle&=&
(2d_{ij}-8)\langle S_{ij}S_{i+1j}\rangle \nonumber\\
&&+\langle S_{i-1j}S_{i+1j}\rangle+\langle S_{ij}S_{i+2j}\rangle
+\langle S_{ij}S_{i+1j+1}\rangle \nonumber\\
&&+\langle S_{ij}S_{i+1j-1}\rangle
+\langle S_{ij-1}S_{i+1j}\rangle+\langle S_{ij+1}S_{i+1j}\rangle \nonumber\\
&&+(1-2P_{ij})\left\{\langle S_{i-1j}S_{ij}S_{i+1j}\rangle
-\frac{3}{2}\langle S_{i+1j}\rangle\right\} \nonumber\\
&&+(1-2P_{i+1j})\left\{\langle S_{ij}S_{i+1j}S_{i+2j}\rangle
- \frac{3}{2}\langle S_{ij}\rangle\right\} \nonumber\\
&&+2(1-2d_{ij}).
\label{2pt:nn}
\end{eqnarray}

In the homogeneous limit $P_{ij}\rightarrow\frac{1}{2}$,
the results in~\cite{krapivsky} are recovered.

\bigskip
\bigskip
\noindent{\bf III. TWO-SITE DISORDER}
\bigskip

We now turn to the case where $P_{ij}$ can take just two different
values, $p$ or $q=1-p$, with an equal number of $p$-sites
and $q$-sites. This corresponds to the model given
in \cite{redner} with equal fluxes of A and B species, 
$\epsilon =|p-\frac{1}{2}|$ and $c_{-}=c_{+}=\frac{1}{2}$, using
the notation given there.

Since $p+q=1$, the global dynamics of the system must be unchanged
under the transformation $(p,q)\rightarrow(1-p,1-q)=(q,p)$.
This symmetry means that the system cannot favour
one state over the other, and so the average of $\langle S_{ij}
\rangle$ taken over the entire $L\times L$ lattice,
$\frac{1}{L^{2}}\sum_{i,j}\langle S_{ij}\rangle$, will always tend to zero
in the $L\rightarrow\infty$ limit.
An important consequence of this is that if a finite system always saturates,
then it does so with equal probability of saturating either to every site
being +1, or every site being -1, and so $\langle S_{ij}
\rangle|_{t=\infty}=0\;\forall i,j$, regardless of whatever $P_{ij}$ may be.
If a reactive steady-state
occurs - that is, if the average saturation time diverges
at least as fast as $e^{L^{2}}$~\cite{ben-A} -
then it should be expected that $\langle S_{ij}\rangle$ may
be non-zero for $t\rightarrow\infty$
(if $p\neq\frac{1}{2}$). It is the purpose of this
section to apply the rate equations derived in section II
to predict the equilibrium value of $\langle S_{ij}\rangle$
on $p$-sites in any such non-trivial steady state.

Although the concentrations of $p$-sites and $q$-sites are equal,
different arrangements of the sites can dramatically alter the
long-time dynamics of the system. For instance, choosing to
split the lattice into two alternating $c(2\times 2)$ sublattices,
with one sublattice full of $p$-sites and the other full of
$q$-sites, results in a system with no non-trivial steady
states for $p\neq 0$ or $1$. Since saturation always occurs,
$\langle S_{ij}\rangle_{t=\infty}=0$ on either type of site.

A more informative model can be constructed by randomly arranging
the $p$ and $q$ sites. This allows for regions of $p$-sites, which will
all tend to be fixed into the same state, and regions
of $q$-sites, which will all tend to be fixed into the other state,
to {\em `pin'} the dynamics into a reactive equilibrium.
Although exact analysis of this model is obviously impossible, a
useful approximation can be made by assuming that every site is
surrounded by exactly 2 $p$-sites and 2 $q$-sites. It is then
possible to write down (\ref{1pt}) and (\ref{2pt:nn}) for the two sorts
of site, $\langle S_{ij}\rangle_{p}$ and $\langle S_{ij}
\rangle_{q}$, and the various two-point functions.

To obtain a closed set of equations, further approximations must be
made to reduce the three-point functions in~(\ref{2pt:nn})
to one and two point functions.
The obvious choice is

\begin{equation}
\langle S_{ij}S_{kl}S_{mn}\rangle\approx \langle S_{ij}S_{kl}\rangle
\langle S_{kl}S_{mn}\rangle,
\end{equation}

\noindent{but this has the unwanted side-effect that $\langle S_{ij}\rangle_{p}
+\langle S_{ij}\rangle_{q}$ $\neq 0$, something which cannot
be true since $p+q=1$. To restore
the required symmetry  we must
also include an alternative three-point approximation,}

\begin{equation}
\langle S_{ij}S_{kl}S_{mn}\rangle\approx \langle S_{ij}\rangle
\langle S_{kl}S_{mn}\rangle.
\end{equation}

For greater clarity, we denote the one-spin correlation function\linebreak
$\langle S_{ij}\rangle_{p}=-\langle S_{ij}\rangle_{q}$ by $y_{p}$,
the two-spin correlation function between two NN
$p$-sites (or, equivalently, two NN $q$-sites) by $z_{pp}$, and 
use $z_{pq}$ for the two-point function between nearest neighbour $p$ and
$q$ sites. Setting $\tau =1$, we can now obtain a closed set of equations,

\begin{eqnarray}
\label{dyp}
2\frac{d}{dt}y_{p}&=&-2y_{p}+(1-2p)(z_{pq}+z_{pp}-2), \\
4\frac{d}{dt}z_{pp}&=&(4pq-8)z_{pp}-3(1-2p)y_{p}+
\nonumber\\
&&\{(1-2p)y_{p}+3z_{pp}\}(z_{pp}+z_{pq})+(2-4pq),\\
4\frac{d}{dt}z_{pq}&=&4pq-(4pq+6)z_{pq}+3(1-2p)y_{p}+3z_{pq}(z_{pp}
+z_{pq}).
\label{dzpq}
\end{eqnarray}

The most constructive way to test the validity of this analysis is to
compare the value of $y_{p}$ at equilibrium, as predicted by
iterating eqns.(\ref{dyp}--\ref{dzpq}), to numerical
simulations. In these simulations,
the sites are initially randomly filled with $+1$'s or $-1$'s, so
the corresponding initial conditions for the iteration procedure are
$y_{p}|_{t=0}=z_{pp}|_{t=0}=z_{pq}|_{t=0}=0$. The results are
compared in fig.1., where the simulation results compare favourably
with the approximate analysis, the agreement improving for larger
values of $p$.

Note that even when $p=1$, $y_{p}$ does
{\em not} tend to +1, either in the theory or in the numerical work.
This is because it is possible to have a jammed state where, for instance,
a $q$-site surrounded by 4 $p$-sites may initially start at +1 but be
unable to change, since if all 4 NN $p$-sites get fixed into a +1
state before they
have reacted with the central $q$-site, then the $q$-site will never be
able to react and so it will stay as +1 for all time, despite the fact
that it has $P_{ij}=0$.
 
\bigskip
\bigskip
\noindent{\bf IV. INHOMOGENEOUS DESORPTION}
\bigskip

We now turn to an enhanced model studied by Fichthorn, Gulari
and Ziff~\cite{ziff}, which introduces noise into the system
in the form of the desorption of A and B species from the substrate.
They demonstrated numerically, later confirmed by mean-field
analysis~\cite{considine}, that even a small desorption rate induces
steady-state reactivity onto finite lattices.
In our version of the model, sites vacated by desorption are
refilled by an A or a B as defined
by the inhomogeneity matrix, which we now call $Q_{ij}$.
$Q_{ij}$ differs from $P_{ij}$ in that now it 
{\em only} applies to sites refilled after desorption - sites vacated
after an \mbox{$A+B\rightarrow 2S$}
reaction have an equal chance of being refilled either by
an A or by a B. Thus, the reaction kinetics alone are the same as the usual
homogeneous model, and the $U_{ij}$,$V_{ij}$ and $W_{ij}$ operators
without the desorption take their simpler form found
by setting $P_{ij}=\frac{1}{2}$ in~(\ref{uij}--\ref{wij}). Explicitly,

\begin{eqnarray}
\label{homo-start}
8\tau_{1}U_{ij}(S)&=&4-S_{ij}(S_{i+1j}+S_{i-1j}+S_{ij+1}+S_{ij-1}),\\
8\tau_{2}V_{ij}(S)&=&1-S_{ij}S_{i+1j},\\
8\tau_{2}W_{ij}(S)&=&1-S_{ij}S_{ij+1}.
\label{homo-end}
\end{eqnarray}

To include inhomogeneous desorption within this formulation,
we replace $U_{ij}$ with $U^{d}_{ij}$,

\begin{equation}
U^{d}_{ij}=U_{ij}+\frac{1}{2\tau_{3}}\{1+S_{ij}(1-2Q_{ij})\},
\label{uij_d}
\end{equation}

\noindent{where, as in~\cite{krapivsky}, we introduce a
renormalised time scale $\tau$ and the spin-flip parameter $\gamma$, defined by
}

\begin{eqnarray}
\frac{1}{\tau}&=&\frac{1}{\tau_{1}}+\frac{1}{\tau_{2}}+\frac{1}{\tau_{3}}, \\
\gamma&=&1-\tau/\tau_{3}.
\end{eqnarray}

The one point spin-correlation rate equation can now be
recalculated using~(\ref{k:master}) and (\ref{homo-start}--\ref{uij_d}),

\begin{equation}
\label{desorb}
4\tau\frac{d}{dt}\langle S_{ij}\rangle = \gamma\Delta_{ij}
\langle S_{ij}\rangle-4(1-\gamma)\{\langle S_{ij}\rangle
+(1-2Q_{ij})\}.
\end{equation}

This can be solved by using a generating function, $G(X,Y,t)$, defined in
terms of the time-dependant one-spin correlation function
$\langle S_{ij}\rangle$,

\begin{equation}
G(X,Y,t)=\sum_{i=-\infty}^{\infty}\sum_{j=-\infty}^{\infty}
X^{i}Y^{j}\langle S_{ij}\rangle.
\label{gen}
\end{equation}

Combining this with~(\ref{desorb}) gives rise to a differential
equation for $G$,

\begin{equation}
\frac{\partial G}{\partial t} = \frac{G}{\tau}\left\{
\frac{\gamma}{4}\left(X+\frac{1}{X}+Y+\frac{1}{Y}\right) -1\right\}
-\frac{1}{\tau_{3}}\sum_{i,j=-\infty}^{\infty}X^{i}Y^{j}(1-2Q_{ij}).
\label{dgdt}
\end{equation}

Noting that, except for G(X,Y,t), the right hand side of~(\ref{dgdt})
is independent
of time, it is not difficult to derive an explicitly time-dependent expression
for $\langle S_{ij}\rangle$ in terms of its initial state,
$\sigma_{ij}=\langle S_{ij}\rangle|_{t=0}$,

\begin{eqnarray}
\langle S_{ij}\rangle &=& e^{-t/\tau}\sum_{k,l=-\infty}^{\infty}
\sigma_{kl}I_{i-k}\left(\frac{\gamma t}{2\tau}\right)
I_{j-l}\left(\frac{\gamma t}{2\tau}\right)
\nonumber\\ &&
-\frac{1}{\tau_{3}}\sum_{k,l=-\infty}^{\infty}(1-2Q_{kl})
\int_{0}^{t}e^{-t'/\tau}
I_{i-k}\left(\frac{\gamma t'}{2\tau}\right)
I_{j-l}\left(\frac{\gamma t'}{2\tau}\right) dt',
\label{desij}
\end{eqnarray}

\noindent{where $I_{i}(t)$ is the $i^{th}$ order modified Bessel function. In
the special case $\Delta_{ij}Q_{ij}=0$ it is possible
to to rewrite the second term on the right hand side of~(\ref{desij}) as}

\begin{equation}
-\frac{1}{4\tau_{3}}\sum_{k,l=-\infty}^{\infty}(1-2Q_{kl})\left\{
f_{i-k+1\:j-l}+f_{i-k-1\:j-l}+f_{i-k\:j-l+1}+f_{i-k\:j-l-1} \right\},
\label{delf}
\end{equation}

\noindent{where for clarity we have introduced}

\begin{equation}
f_{ij}(t)=\int_{0}^{t}e^{-t'/\tau}
I_{i}\left(\frac{\gamma t'}{2\tau}\right)
I_{j}\left(\frac{\gamma t'}{2\tau}\right)
dt',
\end{equation}

\noindent{which obeys the identity}

\begin{eqnarray}
f_{i+1j}+f_{i-1j}+f_{ij+1}+f_{ij-1}&=&\frac{4}{\gamma}f_{ij}
-\frac{4\tau}{\gamma}\delta_{i0}\delta_{j0}
\nonumber\\
&&+\frac{4\tau}{\gamma}e^{-t/\tau}
I_{i}\left(\frac{\gamma t}{2\tau}\right)
I_{j}\left(\frac{\gamma t}{2\tau}\right),
\label{feqn}
\end{eqnarray}

\noindent{with $\delta_{ij}$ the usual Kr\"{o}necker Delta.
Substituting~(\ref{feqn}) into (\ref{desij}) and (\ref{delf}) results in
an exact expression,
}

\begin{equation}
\langle S_{ij}\rangle=2Q_{ij}-1\: +e^{-t/\tau}\sum_{k,l=-\infty}^{\infty}
(1-2Q_{kl}+\sigma_{kl})
I_{i-k}\left(\frac{\gamma t}{2\tau}\right)
I_{j-l}\left(\frac{\gamma t}{2\tau}\right).
\end{equation}

So when $\Delta_{ij}Q_{ij}=0$, $\langle S_{ij}\rangle\rightarrow 2Q_{ij}-1$
exponentially as $t\rightarrow\infty$,
again in agreement with the homogeneous result of~\cite{ziff}.
With desorption, jamming is no longer possible and so now
$\langle S_{ij}\rangle\rightarrow 1$ when $Q_{ij}=1$. Although this final
solution is exact, it is hard to see what physical applications a mixed
homo/inhomogeneous model such as this one may have.

\bigskip
\bigskip
\noindent{\bf V. INHOMOGENEOUS ONE-SPECIES DESORPTION}
\bigskip

Whilst investigating the monomer-dimer model, Ziff, Gulari and
Barshad~\cite{ziff2} briefly discussed the additional feature of allowing
just the monomers to desorb. Physically, this corresponds
to the reaction \mbox{$2CO+O_{2}\rightarrow2CO_{2}$} where only
the $CO$ can desorb from the substrate, which is a good approximation 
for this reaction at the usual operating temperatures.

To apply a similar principle to our monomer-monomer model, we extend
the analysis in section IV to allow for the desorption of A-species
only, with the inhomogeneity matrix $Q_{ij}$ only applying to sites vacated
after desorption. Thus, the flip-exchange operators
are unchanged from~(\ref{homo-start}--\ref{homo-end}),
but now we replace $U_{ij}$ with

\begin{equation}
U^{d}_{ij}=U_{ij}+\frac{1-Q_{ij}}{2\tau_{3}}\left(1+S_{ij}\right).
\end{equation}

Furthermore, $Q_{ij}$ is also taken to be a constant matrix,
$Q_{ij}=q\;\forall i,j$.
The rate equation for the one-spin correlation function~(\ref{desorb})
is now

\begin{equation}
\label{A-desorb}
4\tau\frac{d}{dt}\langle S_{ij}\rangle = \Delta_{ij}
\langle S_{ij}\rangle-\gamma (1-q)(1+\langle S_{ij}\rangle).
\end{equation}

The definitions of $\tau$ and $\gamma$ have now altered from the
previous case,

\begin{eqnarray}
\frac{1}{\tau}&=&\frac{1}{\tau_{1}}+\frac{1}{\tau_{2}}, \\
\gamma &=& \frac{4\tau}{\tau_{3}}.
\end{eqnarray}

Applying the same generating function~(\ref{gen}) results in a new
partial differential equation for $G(X,Y,t)$ acting on an $L\times L$
lattice,

\begin{equation}
\frac{\partial G}{\partial t}=\frac{G}{\tau}\left\{
\frac{1}{4}\left(X+\frac{1}{X}+Y+\frac{1}{Y}\right)-1\right\}
-\frac{1-q}{\tau_{3}}(G+L^{2}).
\end{equation}

Continuing as before, an explicit time-dependent expression
for $\langle S_{ij}\rangle$ is reached,

\begin{eqnarray}
\langle S_{ij}\rangle&=&e^{-t\left(\frac{1}{\tau}+\frac{1-q}{\tau_{3}}\right)}
\sum_{k,l=-L/2}^{L/2}\sigma_{kl}
I_{i-k}\left(\frac{t}{2\tau}\right)
I_{j-l}\left(\frac{t}{2\tau}\right)
\nonumber\\ &&
-\frac{(1-q)L^{2}}{\tau_{3}}\int_{0}^{t}
e^{-t'\left(\frac{1}{\tau}+\frac{1-q}{\tau_{3}}\right)}
I_{i}\left(\frac{t'}{2\tau}\right)
I_{j}\left(\frac{t'}{2\tau}\right) dt'.
\end{eqnarray}

\bigskip
\bigskip
\noindent{\bf VI. CONCLUSIONS AND DISCUSSION}
\bigskip

We have introduced a methodology for dealing with the effects of generalised
surface disorder on the monomer-monomer reaction process
\linebreak \mbox{$A+B\rightarrow2S$},
by mapping the system in the reaction-controlled limit onto an
Ising Model. The two-dimensional rate equations were derived, including
the very concise one-spin correlation equation~(\ref{1pt}), and used to
study the special case of two-site disorder.
Here, it was found that the global system dynamics
are sensitive to the choice of layout of the two different types of site.
Catalysts consisting of two different molecules arranged in a regular manner,
such as on two alternating $c(2\times2)$ sublattices, allow for no
reactive equilibrium and will always saturate on finite lattices.
Choosing to randomly arrange the sites, however, allowing compact
clusters of the same site, was shown to produce a reactive steady-state.
Analysis based on the rate equations was used to predict the concentration
of A's and B's on the different types of site, showing reasonable
agreement between theory and simulation despite the rather crude approximations
involved in the analysis.

The model was then extended to include
desorption from the substrate, either by one or both species, and was
solved exactly in both cases.

Extending this work to dimensions other
than $d=2$ is straightforward once the mapping
onto the Ising model has been achieved. Indeed, the rate equations for
$d=1$ can been immediately seen from those given
here~(\ref{1pt}--\ref{2pt:nn}).
We have focused on $d=2$ since the most useful physical
application is of surface catalysis.

It should be noted that the definition of inhomogeneity we chose to
employ here is only one of many ways of modelling surface disorder.
For instance, requiring that each site be `hit' a different number of
times before adsorbing a particle, or assigning a quenched random
`energy' to each site and always adsorbing the particles onto the
vacant site with the lowest energy, are just two alternative
possibilities. We intend to study some of these in future work. 

\pagebreak

\newpage
Figure 1. Plot of $p$ vs. $y_{p}|_{t=\infty}$. The line gives the values
predicted by the rate equations. Numerical simulation results are plotted
as crosses. The simulations were performed on a $200\times200$ lattice,
and averaged over 100 runs.

\newpage
\begin{figure}
\epsfxsize=\linewidth
\epsffile{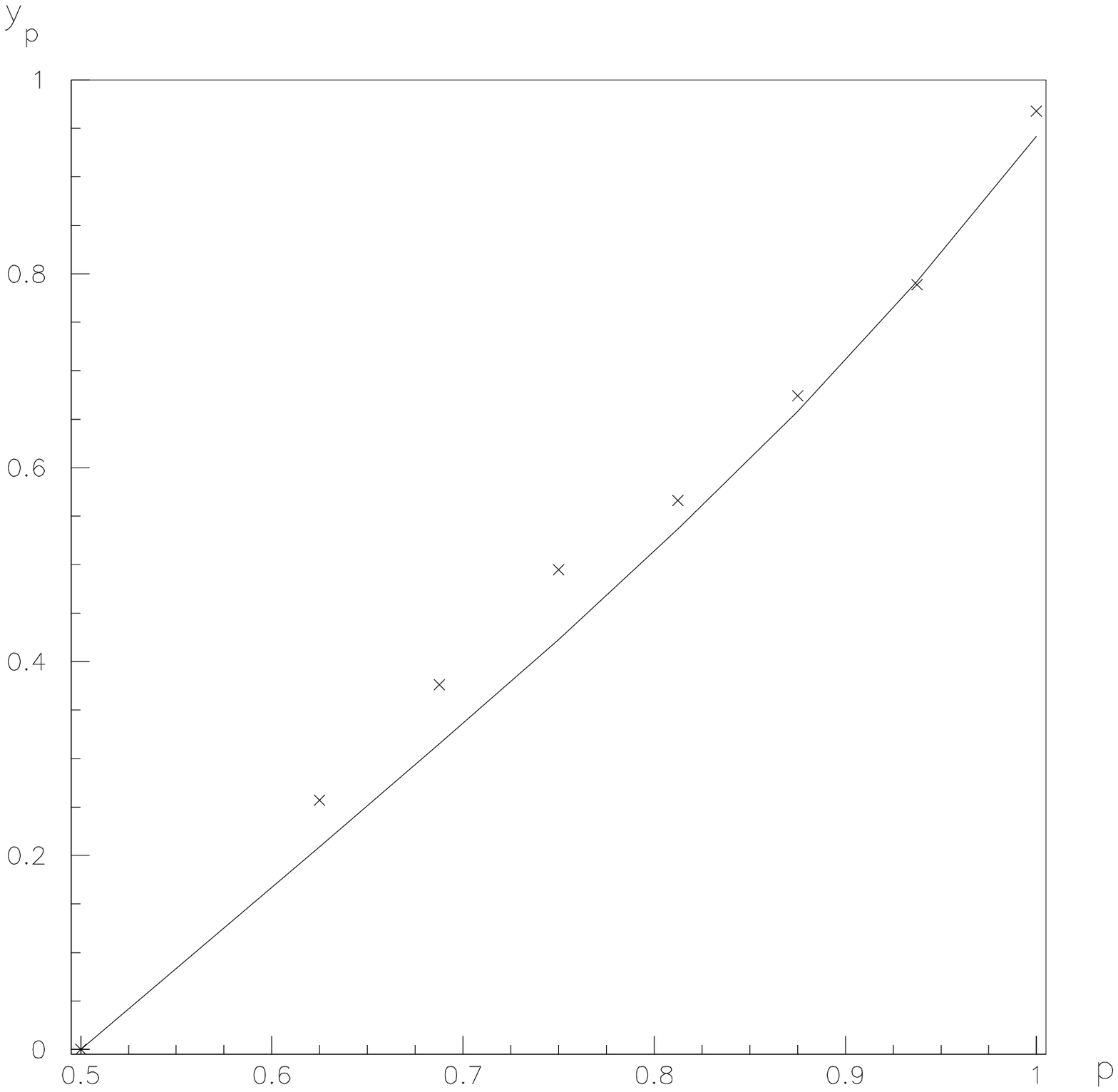}
\end{figure}


\begin{thebibliography}{99}

\bibitem{ziff2} R.Ziff, E.Gulari and Y.Barshad, Phys.Rev.Lett {\bf 56}
2553(1986).

\bibitem{ziff3} R.Ziff and K.Fichthorn, Phys.Rev.B {\bf 34} 2038(1986).

\bibitem{meakin} P.Meakin and D.Scalapino, J.Chem.Phys. {\bf 87}
731(1987).

\bibitem{evans} J.Evans and T.Ray, Phys.Rev.E {\bf 47} 1018(1993).

\bibitem{ben-A} D.ben-Avraham, S.Redner, D.B.Considine and P.Meakin,
J.Phys.A {\bf 23} L613(1990).

\bibitem{krapivsky} P.L.Krapivsky, Phys.Rev.A {\bf 45} 1067(1992).

\bibitem{zhuo} J.Zhuo, S.Redner and H.Park, J.Phys A {\bf 26} 4197(1993).

\bibitem{vlachos} D.Vlachos, L.Schmidt and R.Aris, J.Chem.Phys {\bf 93}
8306(1990).

\bibitem{redner} L.Frachebourg, P.L.Krapivsky and S.Redner,
Phys.Rev.Lett {\bf 75} \newline 2891(1995).

\bibitem{glauber} G.Forgacs, D.Mukamel and R.A.Pelcovits, Phys.Rev.B {\bf 30}
205(1984).

\bibitem{ziff} K.Fichthorn, E.Gulari and R.Ziff, Phys.Rev.Lett {\bf 63}
1527(1989)

\bibitem{considine} D.Considine, S.Redner and H.Takayasu,
Phys.Rev.Lett {\bf 63} 2857(1989).

\end{thebibliography}
\end{document}